\documentclass[12pt]{article}

\usepackage{amsmath}
\usepackage{amssymb}
\usepackage{graphicx}
\usepackage{bm}
\usepackage[top=3cm, bottom=3.25cm, left=2.75cm, right=2.75cm]{geometry}
\usepackage{mathrsfs}
\usepackage{subfig}
\usepackage{cite}
\usepackage{multirow}

\def\be{\begin{equation}}
\def\ee{\end{equation}}
\def\bea{\begin{eqnarray}}

\def\bea{\end{eqnarray}}
\def\eea{\end{eqnarray}}
\def\bs{\begin{split}}
\def\es{\end{split}}
\def\ni{\noindent}

\def\bi{\begin{itemize}}
\def\ei{\end{itemize}}

\def\a{\alpha}

\def\e{\epsilon}
\def\vare{\varepsilon}

\def\d{\delta}

\def\o{\omega}

\def\f{\frac}

\begin{document}

\begin{center}
\LARGE{Influence of Fr\"{o}hlich polaron coupling on renormalized electron bands in polar semiconductors. Results for zincblende GaN}
\end{center}

\vspace{1 mm}
\centerline{\large{Jean Paul Nery \footnote{jeanpaul240@gmail.com} and Philip B. Allen \footnote{philip.allen@stonybrook.edu}}}
\ni \textit{Physics and Astronomy Department, Stony Brook University, Stony Brook, NY 11794-3800, USA}

\vspace{1 mm}
\centerline{July 22, 2016}
\vspace{2 mm}

\ni We develop a simple method to study the zero-point and thermally renormalized electron energy 
$\vare_{\mathbf{k}n}(T)$ for $\mathbf{k}n$ the conduction band minimum or valence maximum in 
polar semiconductors. We use the adiabatic approximation, including an imaginary broadening 
parameter $i\delta$ to supress noise in the density-functional integrations. 
The finite $\delta$ also eliminates the polar divergence which is an artifact of the adiabatic approximation.
Non-adiabatic Fr\"{o}hlich polaron methods then provide analytic expressions for the missing part 
of the contribution of the problematic optical phonon mode. We use this to correct the renormalization 
obtained from the adiabatic approximation. Test calculations are done for zincblende GaN for an 
18$\times$18$\times$18 integration grid. The Fr\"ohlich correction is of order -0.02 eV for the 
zero-point energy shift of the conduction band minimum, and +0.03 eV for the valence band 
maximum; the correction to renormalization of the 3.28 eV gap is -0.05 eV, a significant fraction 
of the total zero point renormalization of -0.15 eV.

\section{Introduction}

Electron quasiparticles in crystals form energy bands $\varepsilon_{\mathbf{k}n}$.  Computations normally use the
Born-Oppenheimer approximation, that atoms are fixed rigidly at crystalline coordinates.  Vibrations
around these fixed coordinates (phonon quasiparticles) are the main cause of temperature-dependent
shifts of the electron bands.  At temperature $T$ the shift is typically $2-4k_B T$, which can have
noticeable effects on electron behavior in semiconductors.  There is also a zero-point shift, caused by
phonon zero-point fluctuations, which is comparable in size to the thermal shift at room temperature.
Cardona and collaborators \cite{Olguin,Thewalt} have given 
brief reviews of the vibrational renormalization of semiconductor bands.
Since energy band calculations omit these effects, a correction should be made when comparing
with experiment.  

These effects have an analog in electronic energy levels in molecules.  When an electron is excited,
interatomic separations and vibrational spectra are altered compared to
the ground state.  To compute the correct electron excitation energy, Born-Oppenheimer energies
are not enough.  This topic is usually described as ``Franck-Condon effects''  
\cite{Franck,Condon,DeVreese,Berry}.  Zero-point vibrational
contributions to a molecular excited state energy are different from the zero-point vibrational contributions
to the ground state energy.  In the molecule, one generally thinks of the change in vibrational energies
caused by electronic excitation, whereas in the crystal one generally thinks of the change in electronic
energies caused by vibrational excitation.  These two points of view are united by what is known \cite{Hui}
as ``Brooks' Theorem'' \cite{Brooks}: the shift in an electron energy $\varepsilon_{\mathbf{k}n}$ caused by a unit 
increase in phonon occupancy of mode $\omega_{\mathbf{q}j}$ equals the shift of the phonon
energy $\hbar\omega_{\mathbf{q}j}$ caused by a unit increase in electron occupancy 
$\varepsilon_{\mathbf{k}n}$.

Computation by density functional theory (DFT) of the temperature dependence of electronic properties of semiconductors and insulators, and also metals, has grown recently \cite{Giustino_Louie_Cohen, Gibbs, Ponce_Gonze, Ponce_Gonze2, Antonius_Gonze, metal1, metal2, Brown}. Fits to experimental data with different models have also been done \cite{Manjon, Cardona_Kremer, Bhosale}. For arbitrary bands, the electron-phonon contribution to the renormalization ($E_{\mathbf{k}n}-\vare_{\mathbf{k}n}$) of the electronic bands, to second order in the ion's displacement, is
 
\begin{flalign}
\begin{aligned}
E_{\mathbf{k}n}-\vare_{\mathbf{k}n} & = \f{1}{N} \sum_{\mathbf{q}jn'}^{\rm BZ} |\langle \mathbf{k}n|H^{(1)}_j|\mathbf{k+q}n'\rangle|^2 \left[ 
\frac{n_{\mathbf{q}j} + 1-f_{\mathbf{k+q}n'}}{\varepsilon_{\mathbf{k}n}-\varepsilon_{\mathbf{k+q}n'}-
\hbar\omega_{\mathbf{q}j}+i \eta}+ \right. \\
& \left. \phantom{=\sum_{\mathbf{q}n'j}^{\rm BZ} |\langle \mathbf{k}n|H^{(1)}_j|\mathbf{k+q}n'\rangle|^2 + +} \f{n_{\mathbf{q}j}+f_{\mathbf{k+q}n'}}{\varepsilon_{\mathbf{k}n}-\varepsilon_{\mathbf{k+q}n'}+
\hbar \omega_{\textbf{q}j}+i \eta} \right] && \\ 
& + \f{1}{N}\sum_{\mathbf{q}j}^{\mathrm{BZ}} \langle \mathbf{k}n | H^{(2)}_{jj}|\mathbf{k}n \rangle [2 n_{\mathbf{q}j} +1] &&
\label{eq:general}
\end{aligned}
\end{flalign}

\ni Here, $\langle \mathbf{k}n|H^{(1)}_j|\mathbf{k+q}n'\rangle$  is the matrix element for scattering an electron $\mathbf{k}$ by a phonon $\mathbf{q}$; it has units of energy and a typical size of roughly the geometric mean of electron and phonon energies. The Debye-Waller term $ \langle \mathbf{k}n | H^{(2)}_{jj}|\mathbf{k}n \rangle$ is the second order interaction energy involving two phonons $\mathbf{q}j$ and $\mathbf{q}'j'$, but only $\mathbf{q}j = -\mathbf{q'}j'$ enters in lowest order. The Fermi-Dirac and Bose-Einstein equilibrium occupation factors are denoted $f$ and $n$.  The infinitesimal parameter $i \eta$ ensures the real and imaginary parts are well defined. Only the real part is discussed here.  We omit the smaller thermal expansion contribution in this work.
 
The formulas used by Allen, Heine and Cardona \cite{Allen_Heine, Allen_Cardona} intentionally drop the phonon energy $\pm\hbar\omega_{\mathbf{q}j}$ from the denominators in comparison with the electron energy difference $\varepsilon_{\mathbf{k}n} -\varepsilon_{\mathbf{k+q}n'}$. This is an adiabatic approximation. The justification is that, in semiconductors, typical energy denominators are much larger than $\hbar\omega_{\mathbf{q}j}$. However, it was pointed out by Ponc\'e {\it et al.} \cite{Ponce_Gonze} that for polar materials, it is necessary to keep the $\pm\hbar\omega_{LO}$ for longitudinal optic (LO) modes to avoid an unphysical divergence in the intraband ($n^\prime=n$) term at band extrema, 
caused by the adiabatic treatment of the long-range Fr\"ohlich-type electron-phonon interaction.

A converged non-adiabatic evaluation of Eq. \eqref{eq:general}, summed on a fine enough mesh to
accurately get the Fr\"ohlich part of the renormalization, requires a very fine and very expensive
mesh.  Our aim in this paper is to explore a simplified method that works adequately on a coarser
mesh.  We test our method by computations for zincblende (cubic) GaN, abbreviated c-GaN. 
Our corrections use the effective mass approximation ($\vare_{\textbf{k}} = \hbar^2 k^2/2m^*$ for
band edges near $\textbf{k}=0$.  This works well for the conduction band where $m^\ast$ is
small, $\approx 0.16 m_e$.  The top of the valence band is triply-degenerate 
(because we ignore the spin-orbit interaction) and involves higher effective masses, which work
a bit less well. 

One reason for choosing Gallium nitride is its useful properties, including a high thermal conductivity 
\cite{Slack}, and a high melting point that allows it to operate at high temperatures. 
Its wide and direct band gap make it efficient for lasers \cite{Nakamura}, and for 
high-power and high-frequency electronic devices \cite{Khan, Mohammad, Chung}. 
It is used in white LED's. 
Alloying with InN and AlN allows engineering of optical and electrical properties \cite{Strite}. 
For simplicity we study c-GaN rather than the more stable wurtzite (hexagonal) GaN, or h-GaN.
Although h-GaN has been more thoroughly studied, c-GaN has several advantages: 
it has better n and p doping properties \cite{Brandt, Lin}, higher saturated electron drift 
mobilities \cite{Mullhauser, Strite}, and it is convenient to work in the 510 nm region.

\section{``Adiabatic plus $i\delta$'' approximation corrected using effective-mass theory}

For convenience, we assume (correctly for c-GaN) that band extrema are at $\mathbf{k}=0$. 
The Fr\"ohlich part of the integral in Eq. (\ref{eq:general})
involves $\int d^3q$ and a factor $1/q^2$ from the long-range polar electron-phonon matrix element.  
If the $\pm\hbar\omega_{\rm LO}$ is omitted, then in the small-$\mathbf{q}$ Fr\"ohlich region,
denominators in Eq. (\ref{eq:general}) behave as $q^2$.  The
integral then involves $\int dq/q^2$ which diverges at $q=0$.  
When $\pm\hbar\omega_{\rm LO}$ is kept, the divergence
is removed from the first denominator in Eq. (\ref{eq:general}), and the singularity in the second
denominator is integrable.  When ${\mathbf k}$ is not chosen to be 0, 
there are (integrable) singular denominators 
$\varepsilon_{\mathbf{k}n} - \varepsilon_{\mathbf{k}+\mathbf{q}n'}\pm\hbar\omega_{\mathbf{q}j}
\rightarrow 0$ on extended surfaces in $\mathbf{q}$-space.  
All these cases create problems if integrated numerically by summing points on a simple mesh.
Regardless of how dense the $\textbf{q}$-mesh is, singular integrals of this type 
do not converge (as already noted in \cite{Allen_Cardona}) except with a carefully tempered
mesh, designed to give the correct principal-value treatment in three dimensions. 
A useful procedure is to change the $i \eta$ in the denominator to a finite imaginary energy $i\delta$. 
Convergence in this parameter was studied by Ponc\'e {\it et al.}
\cite{Ponce_Gonze}. Since the true result around this type of singularity 
integrates to a small contribution when done correctly, it is safe  
to add a finite imaginary energy consistent with the mesh size.  Unlike Ponc\'e {\it et al.},
we do not need $\delta$ to be particularly small or less than $\hbar\omega_{\rm LO}$.
Specifically, $\delta$ should not be smaller than the typical energy jump
$\Delta_s\varepsilon=\vare_{\mathbf{k}+\mathbf{q}+\Delta\mathbf{q}}-\vare_{\mathbf{k}+\mathbf{q}}$
associated with the mesh size $\Delta\mathbf{q}$ when $\vare_{\mathbf{k}+\mathbf{q}}$
lies near the singularity surface.  The singular part of the integrand $1/\Delta_s\varepsilon$
is then replaced by $\Delta_s\varepsilon/(\Delta_s\varepsilon^2 +\delta^2)$.  
The subscript ``s'' indicates ``singularity.''  Errors associated
with the random location of mesh points relative to the singularity surface are then reduced
from $N_s^{1/2}/\Delta_s\varepsilon$ to $N_s^{1/2}\Delta_s\varepsilon/\delta^2$, where
$N_s$ is the number of mesh points neighboring the singularity surface.  When the
singularity is at $\mathbf{k}=0$, $N_s \approx 1$, but for an extended singularity, the value
of $N_s$ is likely to be of order $N_{\rm mesh}^{2/3}$.  Therefore the value of
$\delta$ should be greater than $\Delta_s\epsilon$ or $N_{\rm mesh}^{1/6}\Delta_s\epsilon$,
depending on whether the singularity is at a point or on an extended area in $\mathbf{k}$-space.
At a minimum or maximum (local or absolute) of $\vare_{\mathbf{k}n}$, there is a singular point
which requires a non-adiabatic treatment in polar materials.  When $\vare_{\mathbf{k}n}$
is not at an absolute band maximum or minimum, the extended singularity surface can be
safely approximated by replacing $\pm \omega_{\mathbf{q}j}$ by $i\delta$.  The reason is, if the surface is
redefined by $\Delta\vare=0$ instead of $\Delta\vare\pm \omega_{\mathbf{q}j}=0$, it causes only a small
shift of the surface in $\mathbf{k}$-space.  This should do little to change the small remainder
after principal-parts cancellation of the singularity.  The replacement of $\pm\omega_{\mathbf{q}j}$
by $i\delta$ is what we call the ``adiabatic + $i\delta$'' method.

When the state of interest $\mathbf{k}n$ is a (local or absolute) band extremum 
(taken here to be $\mathbf{k}=0$),
replacement of $\pm\hbar\omega_{\rm LO}$ by a finite $i\delta$  does not
correctly treat the Fr\"ohlich intraband renormalization effect.  
This is especially true in the first denominator
of Eq. (\ref{eq:general}).  This ``emission term'' with $n_{\mathbf{q}j}+1$ in the numerator,
integrates only over one side of the singularity, and thus has no principal-parts cancellation.  
The long-range polar
interaction, when treated correctly (non-adiabatically), makes an additional renormalization.  
Our aim is to use a mesh fine enough to
capture all the less singular contributions, but coarse enough for rapid computation 
(for example 20$\times$20$\times$20).  Then to include the Fr\"ohlich
effect, we want to focus on a small $\mathbf{q}$ ``central region'' and treat it by an 
analytic integration using effective mass theory.
For this purpose we need a central region large enough that outside it, 
$\pm\hbar\omega_{\rm LO}$ can be 
safely replaced by $i\delta$, but small enough that inside, the energy 
$\vare_{\mathbf{k},n}$ can be replaced by
$\vare_{0n}+\hbar^2 k^2/2m^\ast$.  The mesh should be fine enough 
that the ``adiabatic plus $i\delta$'' calculation (by mesh summation) is
reasonably converged in the central region, and therefore adequately approximated 
by an analytic effective-mass integration of the ``adiabatic plus $i\delta$'' intraband central region sum. 
If these conditions can be satisfied, then we can subtract
the analytic effective-mass version of the ``adiabatic plus $i\delta$ and add the analytic 
effective-mass version of
the Fr\"ohlich renormalization to get a good computation of the full non-adiabatic theory.

For the direct $\mathbf{k}=0$ gaps of c-GaN (the case we study in detail), the relevant energy jump is
$\Delta_s\varepsilon=(\hbar^2/2 m^\ast) (\Delta q)^2 $, where $\Delta q$ is the size of the $\mathbf{q}$-grid.  
The value of $m^\ast$ for the conduction band is 0.16$m_e$, and
 $\hbar\omega_{\rm LO}$ is 0.089 eV.  A desirable value of $\delta $ is 0.1eV, 
which requires $\Delta q = 0.065 \mathring{A}$ to make $\Delta_s\varepsilon<\delta$.  However, we 
find that an 18$\times$18$\times$18 mesh is sufficient.  This
corresponds to $\Delta q = 0.155 \mathring{A}$.  The reason why this works is because
the grid and the singular point are both centered at $\mathbf{k} =0$.
The integrand is then sampled at symmetric points, an appropriate ``tempered mesh''
that converges with far less noise to the correct principal value integral.
A confirmation that this works comes from the plots of Ponc\'e {\it et al.}
\cite{Ponce_Gonze}.  See for example the middle graph of Fig.6(a), which shows very good convergence
for a 20$\times$20$\times$20 grid.

 \section{Correction formulas}

The full theory is contained in the perturbative expressions worked out by Vogl \cite{Vogl}. The singular part corresponds to the Fr\"{o}hlich polaron \cite{Frohlich}. In \cite{Verdi_Giustino, Mauri} a Vogl expression is studied from an \textit{ab-initio} perspective, and is shown to coincide for small $\mathbf{q}$ with DFPT calculations. All agree that the polaron is the dominant contribution in the small $\textbf{q}$ region, and needs to be treated carefully.

A polaron describes the coupled system of an electron and phonons. Most often, only zero temperature is considered, but the concept works also at $T>0$. The most famous case is the Fr\"ohlich, or ``large'' polaron, present in ionic crystals and polar semiconductors \cite{Frohlich}. Fr\"ohlich theory is designed for the bottom of the conduction band where an effective mass approximation $\vare_{\mathbf{k}} = \hbar^2 k^2/2m^*$ is accurate, and for intraband ($n=n'$) coupling only to the polar LO mode, where $\omega_{\rm LO}$ has negligible $\mathbf{q}$-dependence. It can also be used for the valence band, which will be discussed later. In the conduction band case, the matrix element $|M|^2$ is $4\pi\alpha(\hbar\omega_{\rm LO})^2 (a_{\rm LO}/\Omega_0 q^2)$. It is the factor $q^{-2}$ which comes from long range polarization. The distance $a_{\rm LO}=\sqrt{\hbar/2m^\ast \omega_{\rm LO}}$ is of order 10$\mathring{A}$, larger than
the zero-point root mean square vibrational displacement $u_{\rm LO}=\sqrt{\hbar/2 M_{\textrm{red}} \omega_{\rm LO}}$ by the large factor $\sqrt{M_{\textrm{red}}/m^*}$, where $M_{\textrm{red}}$ is the appropriate ionic reduced mass.  The Fr\"ohlich coupling constant is $\alpha = V_c/\hbar\omega_{\rm LO}$, where $V_c$ is a Coulomb interaction strength $V_c=e^2/(8\pi\tilde{\epsilon}_0 \epsilon^\ast a_{\rm LO})$.  The $\tilde{\epsilon}_0$ is the permittivity of free space, and the $\epsilon^\ast$ is defined in terms of the low and high frequency dielectric constants as $1/\epsilon^\ast=1/\epsilon_\infty - 1/\epsilon_0$. Since we are interested in the renormalization of the band gap, we focus on the band extrema at $\textbf{k} = 0$. For a non-degenerate band ({\it e.g.} the conduction band), the Fr\"{o}hlich contribution 
to the renormalization at temperature $T$ is (see \cite{Callaway} for the $T=0$ result)
\begin{flalign}
\begin{aligned}
\phantom{.} [ E_{\mathrm{k}c} - \vare_{\mathrm{k}c} ]_{\mathrm{Fr},\mathbf{k}=0} = &-\frac{\alpha\hbar  \omega_{LO}}{2\pi^2 a_{LO}}
 \int_0^{q_F} \frac{4\pi q^2 dq}{q^2} 
 \bigg[ \frac{n_B(T)+1}{q^2+a_{LO}^{-2}} + \frac{n_B(T)}{q^2-a_{LO}^{-2}}  \bigg] \\
 =&-\alpha\hbar\omega_{LO} \left\{\frac{\tan^{-1}(q_F a_{LO})}{\pi/2}[n_B(T)+1]+
\f{1}{\pi} \mathrm{ln} \left| \f{q_F-a^{-1}_{LO}}{q_F+a^{-1}_{LO}}\right| [n_B(T)]\right\} 
\label{frohlich}
\end{aligned}
\end{flalign}
where $\o_{\rm LO}$ is the longitudinal optical frequency, and $n_B(T)=1 / [\mathrm{exp}(\hbar \o_{LO}/k T)-1]$ 
is the Bose- Einstein distribution. The radius of integration is $q_F$. 
This and other radii in reciprocal space used in this work, together with their approximate values, 
are included in Table \ref{table}. In most polaron studies, the approximation $q_F\rightarrow\infty$ 
is used.  One might instead use  the radius $q_D$ of the Debye sphere whose volume is the BZ volume. 
However, the integrand becomes inaccurate if $q_F$ is larger than the radius $q_m$ where
the effective mass treatment works well.
The first term of Eq.(\ref{frohlich}) corresponds to phonon emission.
It is included in Fr\"{o}hlich's treatment at $T=0$. 
The second term is only present at non-zero temperature and it corresponds to phonon absorption. 
At $T=0$, extending the sum over the Brillouin Zone to infinity, the famous result \cite{Callaway} is
$E_{\mathbf{k}=0,c}-\vare_{ \mathbf{k}=0,c}=-\alpha\hbar\omega_{\rm LO}$.

In the adiabatic approximation, the term in brackets [ ] in Eq. \eqref{frohlich} 
is replaced by $(2 n_B + 1)/(q^2 - i 2m^* \d / \hbar^2)$, and then the real part is taken:
\begin{equation}
\begin{split}
\phantom{.} [ E_{\mathrm{k}c} - \vare_{\mathrm{k}c} ]_{\mathrm{Ad},\mathrm{k}=0} = &-\frac{\alpha\hbar  \omega_{LO}}{2\pi^2 a_{LO}}
 \Re \int_0^{q_F} \frac{4\pi q^2 dq}{q^2} 
 \left[ \frac{2n_B(T)+1}{q^2-i  2 m^\ast \d /\hbar^2} \right] \\
 = & -\f{\a \hbar \o_{LO}}{a_{LO}} \Re \f{1}{ \pi z} \mathrm{ln} \left(-\f{q_F-z}{q_F+z} \right) [2n_B(T)+1]
\end{split}
\label{adiabatic}
\end{equation}
where $z= \sqrt{ 2 m^\ast \d / \hbar^2} \mathrm{exp}(i \pi/4)$. Subtracting this term from the \textit{ab-initio} calculation and adding the correct Fr\"{o}hlich contribution \eqref{frohlich}, with an appropriate radius of integration $q_c$, we obtain in principle our desired correction.

In the adiabatic approximation, the denominator (for $\mathbf{k=0}$ and $\mathbf{q} \rightarrow 0$) 
is $i \d$, i.e., pure imaginary. Because the energy renormalization is given by the real part, 
the central mesh-cell contribution is 0 in the adiabatic approximation. 
This mis-represents a converged adiabatic calculation (like Eq.(\ref{adiabatic})).
We should not subtract the part of Eq.(\ref{adiabatic}) that represents the missing
contribution from the central grid cell.

To determine the optimal integation radius $q_c$ to use for the correction, we calculate 
the difference between the Fr\"{o}hlich-polaron contribution Eq. \eqref{frohlich} and the 
adiabatic approximation Eq. \eqref{adiabatic} (replacing $\o_{LO}$ with $i 0.1$ eV) for 
different radii $q_F$ of integration. We denote $q_c$ the radius for which the curves differ 
by less than 1 meV for all temperatures, and we refer to it as the convergence radius. 
The adiabatic expression is a good approximation to the Fr\"{o}hlich polaron for radii 
greater than $q_c$. The analysis can be separated in two cases:

(i) $q_{\textrm{mesh}} < q_c$.  This is the case in our calculation, 
for both the conduction and valence band. It is discussed in the Appendix.

(ii) $q_{\textrm{mesh}} \geqslant q_c$.  We will illustrate this case with our c-GaN calculation, 
although the expression in the Appendix is required for a more precise result. 
Since the adiabatic DFT calculation has no contribution from the central cell, 
it does not  have to be subtracted. Therefore, the correction is just given by \eqref{frohlich} 
with $q_F= q_{\textrm{mesh}}$, which is a good enough radius of integration 
since $q_{\textrm{mesh}} \geqslant q_c$. Case (i) is similarly simple, 
but the correction involves an extra term.

As long as the effective mass approximation is accurate, both Eq.(\ref{frohlich}) and Eq.(\ref{adiabatic}),
and thus $q_c$, will be accurate.  If there are worries about the applicability of the effective mass
approximation, then one coud use the small $\mathbf{q}$, intraband, LO phonon 
part of  Eq. \eqref{eq:general} with $\omega_{\rm LO} \rightarrow i \delta$ 
to subtract the adiabatic contribution, without using the effective mass approximation.  
However, this is not necessary for the accuracy of a few meV we are interested in. 

We first study how the correction changes with the integration radius. 
Then we calculate the temperature dependence of the minimum of the conduction band, 
including the polaron correction.  Finally, we study the valence band.  We use ABINIT \cite{Abinit1, Abinit2} to carry out the \textit{ab-initio} calculations.

\section{Results and Discussion}

We use Troullier-Martins pseudopotentials for both Gallium and Nitrogen, 
in the Perdew-Wang \cite{Perdew_Wang} parameterization of local density approximation (LDA), 
generated using the fhi98PP code \cite{fhi98PP}. The Ga-3d electrons are included 
as valence electrons.  We use a 6$\times$6$\times6$ Monkhorst-Pack (MP) \cite{MP} centered \textbf{k}-point grid 
in our calculations, and a high energy cutoff of 1700 eV in order to converge the total energy 
to less than 0.018 meV per atom (h-GaN converges well with 1400 eV).  
The resulting lattice constant is $a = 4.499 \mathring{A}$.  Experimental values are 
4.507 $\mathring{A}$ \cite{Feneberg} and 4.52 $\mathring{A}$ \cite{Bougrov}. 
The phonons and electron-phonon interaction matrix elements use DFPT 
in the rigid-ion approximation, as specified by ABINIT, to speed up calculations. 
It has been shown reliable for simple crystals \cite{Antonius_Gonze}. 
We use an 18$\times$18$\times18$ MP \textbf{q}-point grid, and the adiabatic + i$\delta$
approximation, with $\delta=0.1$ eV.

\subsection{Conduction band}

\begin{figure}
\centering
	 \includegraphics[width=0.8\textwidth]{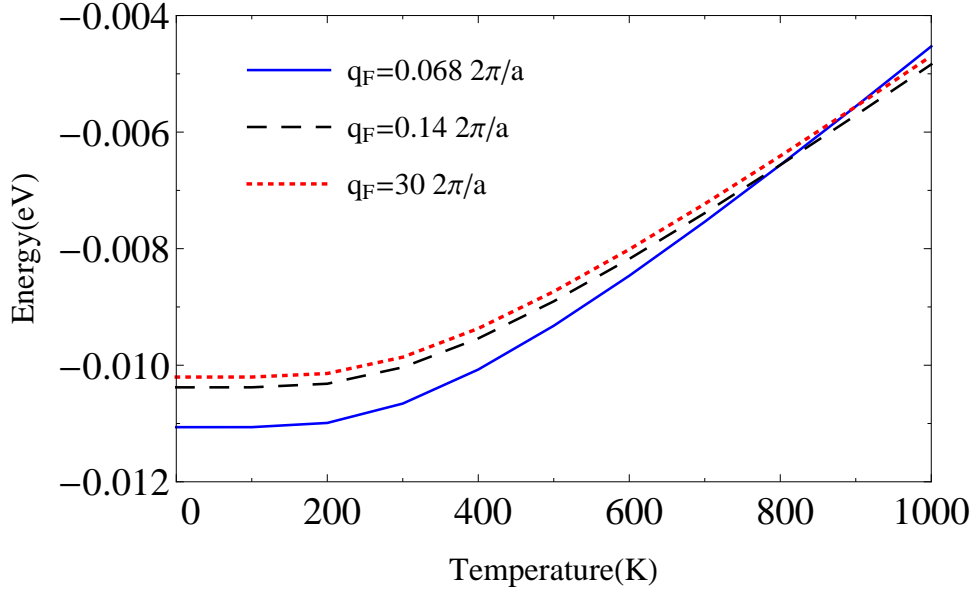}
\caption{Difference between the Fr\"{o}hlich contribution and the corresponding adiabatic approximation with $i \d = i 0.1$eV for the conduction band, for different radii $q_F$ of integration. \label{correction_convergence}}
\end{figure}

The conduction band is very isotropic, with an effective mass $m^*=0.16 m_e$. The differences 
between the Fr\"ohlich contribution and the adiabatic approximation are shown in 
Fig. \ref{correction_convergence}.  Going beyond a radius of integration of $q_F =0.068$ $2 \pi /a$, 
the curves differ by less than 1 meV for all temperatures. Therefore, $q_c = 0.068$ $2 \pi /a$. 
To obtain an accurate result, the analytic integration in \eqref{frohlich} should be restricted 
to a small radius $q_c$ close to $\mathbf{q}=0$, because the effective mass approximation 
is only valid close to $\mathbf{q}=0$. From Table \ref{table}, we see that for the conduction  
band $q_{m} = q_c$, so the method is indeed accurate.

\begin{table}[]
\centering
\begin{tabular}{c|c|cc}
Symbol                        & Definition                                                                                                                                                 & \multicolumn{2}{c}{Approximate choices} \\ \hline
$q_F$                         & Upper limit of Fr\"ohlich integral Eq. \eqref{frohlich}                                                                                                        & \multicolumn{2}{c}{$\f{4 \pi}{3} q_F^3 = \Omega_{BZ}$ }             \\ \hline
\multirow{2}{*}{$q_c$}        & \multirow{2}{*}{\begin{tabular}[c]{@{}c@{}}Convergence radius beyond \\ which $\Delta \vare_{\textrm{Fr}} \approx \Delta \vare_{\textrm{Ad}}$\end{tabular}} & Conduction        & 1.2$q_{\textrm{mesh}}$         \\
                              &                                                                                                                                                            & Valence           & 2.5$q_{\textrm{mesh}}(T=0)$        \\ \hline
\multirow{2}{*}{$q_{m}$} & \multirow{2}{*}{\begin{tabular}[c]{@{}c@{}}Wavevector limit for effective\\ mass approximation\end{tabular}}                                               & Conduction        & 1.2$q_{\textrm{mesh}}$        \\
                              &                                                                                                                                                            & Valence           & 2.5$q_{\textrm{mesh}}$        \\ \hline
$q_{\textrm{mesh}}$                    & $\f{4 \pi}{3}q_{\textrm{mesh}}^3=\f{\Omega_{BZ}}{N}$                                                                                                                     & \multicolumn{2}{c}{$N=18\times18\times18$}         
\end{tabular}
\caption{Definitions and approximate values of the different radii in momentum space used in this work. 
The convergence radius $q_c$ determines the region in which the correction has to be applied. 
Note the similarity between $q_c$ and $q_m^{\ast}$ both for the valence and conduction bands. 
However, $q_c= 6.3 q_{\textrm{mesh}}$ at $T=1000$K because of the absorption term in the valence band. See the discussion in the Appendix.
\label{table}}
\end{table}

\begin{figure}
\centering
	 \includegraphics[width=0.8\textwidth]{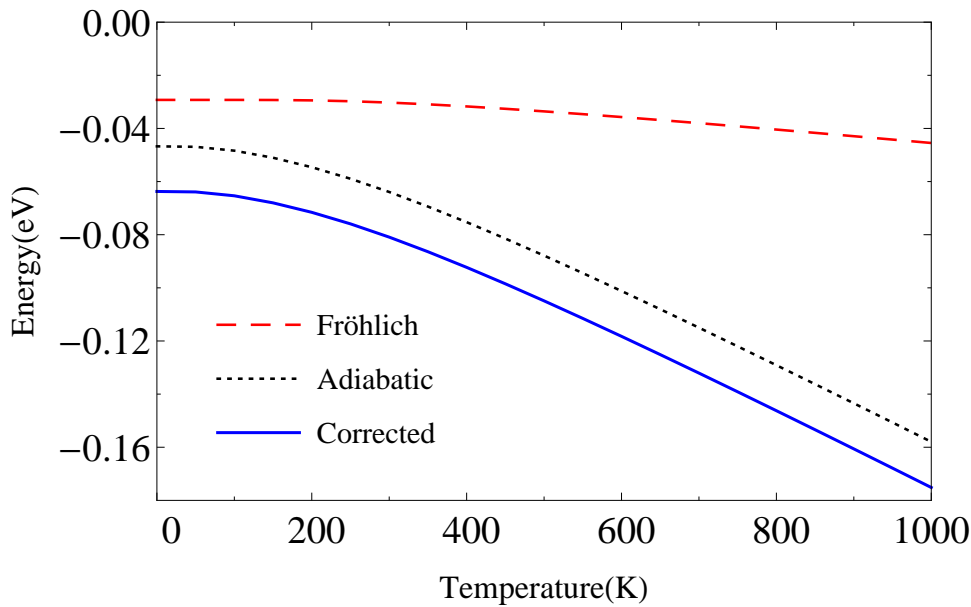}
\caption{Temperature dependence of the conduction band: direct adiabatic calculation (including interband parts in Eq. \eqref{eq:general} in a 18$\times$18$\times$18 MP grid with $\d = 0.1$eV (dotted), corrected calculation (full), and the pure Fr\"{o}hlich term at finite temperature (dashed).\label{adiabatic_corrected}}
\end{figure}

Using the method of case (ii), the corrections at $T=0$ K and $T=1000$ K are $-19$ meV 
and $-22$ meV, respectively. For the more precise method (i) described in the Appendix, 
since actually $q_{\textrm{mesh}} < q_c$, the corrections are -17 meV at $T = 0$ K 
and -17 meV at T = 1000 K. Fig. \ref{adiabatic_corrected} shows the adiabatic calculation of 
$E_{\Gamma c}-\vare_{\Gamma c}$ done with ABINIT, the corrected result, and the total 
Fr\"{o}hlich contribution at finite temperature (taking as $q_F$ the radius of the BZ).

\subsection{Valence band}

For the valence band, the correction is more complicated because of two factors: bands are degenerate and they are not isotropic. Since we are not considering spin-orbit coupling, the top of the valence band is triply degenerate. The $\mathbf{k}\cdot\mathbf{p}$ method fixes the valence band energy dispersion to be the eigenvalues of \cite{Baldereschi}
\be
D=
\begin{bmatrix}
    A k_z^2 + B(k_y^2 + k_z^2)  & C k_x k_y & C k_x k_z \\
    C k_x k_y       & A k_y^2 + B(k_x^2+k_z^2) & C k_y k_z \\
    C k_x k_z       & C k_y k_z & A k_z^2 + B(k_x^2+k_y^2)
\end{bmatrix}
\label{kp_method}
\ee
Comparing with the \textit{ab-initio} calculation, we obtain $A=-3.14\hbar^2/m_e$, $B=-0.61\hbar^2/m_e$ 
and $C =-3.49\hbar^2/m_e$. These correspond, for example, to effective masses $m^\ast = 0.16 m_e$ 
and $0.82 m_e$ in the (100) direction.

Renormalization does not lift the triple degeneracy of the top of the valence band. 
For degenerate and isotropic bands, Trebin and R\"{o}ssler \cite{Trebin} use 
the $\mathbf{k}\cdot\mathbf{p}$ method to generalize Fr\"{o}hlich's result (giving analytic 
expressions). Following their procedure, we write the band renormalization in the 
 case, without requiring band isotropy:
\begin{flalign}
\begin{aligned}
\phantom{.} [E_{\mathbf{k}v}-\vare_{\mathbf{k}v}]_{\mathrm{Fr},\mathbf{k}=0}= \f{e^2}{4 \pi \tilde{\e_0} N \Omega_0} & \f{2 \pi  \hbar \o_{LO}}{\e^\ast} \sum_q^{q < q_F} \sum_{s=1}^3 \f{1}{q^2}|\langle n_s(\mathbf{q}) | n \rangle |^2 \times \\
& \times \Re \left[ \f{n_B(T)+1}{\vare_{\Gamma v}-\vare_{\mathbf{q}n_s}+\hbar \o_{LO}} + \f{n_B(T)}{\vare_{\Gamma v}-\vare_{\mathbf{q}n_s}-\hbar \o_{LO} + i \Delta} \right]
\end{aligned}
\label{frohlich_valence}
\end{flalign}
where $n_s$ indicates the degenerate bands and $|n_s(\mathbf{q}) \rangle$ are the eigenstates 
of Eq.(\eqref{kp_method}) at $\textbf{q}$. The initial state $|n \rangle$ can be any of 
the $\mathbf{k}=0$ degenerate eignestates; all give the same answer. We include a 
small $i \Delta = i 0.001$eV only in the second denominator to allow a good numerical 
evaluation of the principal part. Now, $\vare_{\Gamma}-\vare_{\mathbf{q}} >0$ and the 
factors $n_B+1-f$ and $n_B + f$ have become $n_B$ and $n_B + 1$ instead of $n_B +1$ 
and $n_B$, respectively (because $f_v=1$). As a result, we get an extra minus sign with 
respect to the conduction band; the band renormalization is now positive.

\begin{figure}
\centering
	 \includegraphics[width=0.8\textwidth]{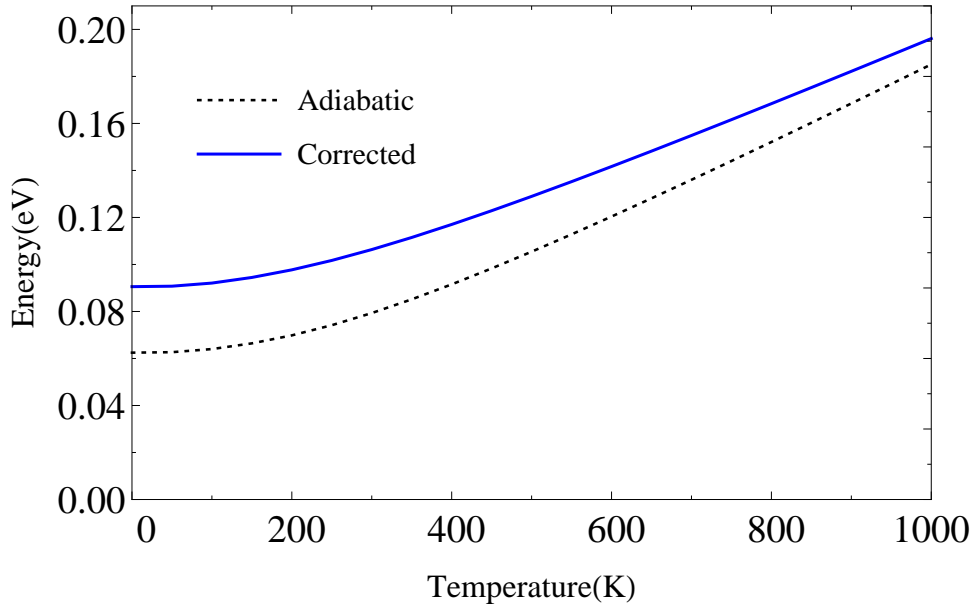}
\caption{Temperature dependence of the valence band: direct adiabatic calculation 
with an 18$\times$18$\times18$ MP grid with $\d = 0.1$eV (dotted) and the corrected 
calculation (full).\label{adiabatic_corrected_val}}
\end{figure}

The adiabatic \textit{ab-initio} calculation gives a band renormalization of 62 meV at T=0 K 
and 185 meV at T=1000 K. The valence band has in case (i) ($q_{\textrm{mesh}}<q_c$) and 
the method is described in the Appendix. The correction is 28 meV and 11 meV at $T=0$ K 
and $T=1000$ K, respectively. Fig. \ref{adiabatic_corrected_val} shows the results for the 
valence band analogous to Fig. \ref{adiabatic_corrected}.



Therefore, we see that the Fr\"ohlich correction provides approximately a constant shift of the 
renormalization by -20 meV for all temperatures in the conduction band. For the valence band, 
it is +28 meV at $T=0$ and it decreases to 11 meV at $T=1000$ K. The Fr\"ohlich correction 
is about 30\% of the total ZPR of both the conduction and valence band. At 1000 K, the 
corrections are between 6-9\% of the total renormalization. 

\section{Conclusions}

Our procedure allows a calculation of the whole electronic renormalization of a polar material,
using the adiabatic approximation with an $i \d$, and an affordable mesh.  The $i \d$ cures the 
divergence of the adiabatic approximation at the extrema of polar materials, but does not correctly
include Fr\"ohlich polaron corrections.
Then we add the Fr\"{o}hlich polaron contribution in the central mesh-cell, 
omitted in the DFT adiabatic calculation due to the pure imaginary denominator $i \d$.
Finally, we add the difference between the Fr\"ohlich and adiabatic expressions if $q_c>q_{\rm mesh}$. 
This method is then a combination of the adiabatic and non-adiabatic approximations. 
We avoid using a very dense $\textbf{q}$-grid by treating the Fr\"ohlich polaron analytically. 
By this method, we calculated for c-GaN the electron-phonon temperature dependence of the 
minimum of the conduction band and the maximum of the valence mand. 
At high temperatures, the method is approximate for the valence band. 
The correction is a significant fraction of the total electron-phonon renormalization, 
although it decreases as the temperature increases.

\section{Acknowledgments}
We thank the Brookhaven National Laboratory Center for Functional Nanomaterials (CFN) under project 33862
for time on their computer cluster. This research also used computational resources at the Stony Brook University
Institute for Advanced Computational Science (IACS). Work at Stony Brook was supported by US DOE Grant No.
DE-FG02-08ER46550. JPN is deeply grateful to Elena Hirsch and the Fundaci\'on Bunge y Born for their financial support during his Master's degree at SBU. JPN also thanks Samuel Ponc\'e for help.

\section{Appendix}

\subsection{Method, case (i)}
Here we describe the method for case (i), where $q_{\textrm{mesh}} < q_c$. The difference between the Fr\"ohlich contribution and the adiabatic expression is small beyond a radius $q_c$. Because of the $i \d$ there is no adiabatic contribution from the central cell (in c-GaN, a truncated octahedron), which can be approximated by a sphere of radius $q_{\textrm{mesh}}$. Instead of subtracting the adiabatic contribution from 0 to $q_c$, we have to subtract it from $q_{\textrm{mesh}}$ to $q_c$. Therefore we need to determine what is the adiabatic contribution in this region.

\subsubsection{Conduction band}

The correction is given by
\begin{flalign}
\begin{aligned}
\Delta(E_{\mathbf{k}c}-\vare_{\mathbf{k}vc})_{\mathbf{k}=0} =&-\alpha\hbar\omega_{LO} \left\{\frac{\tan^{-1}(q_c a_{LO})}{\pi/2}[n_B(T)+1]+
\f{1}{\pi} \mathrm{ln} \left| \f{q_c-a^{-1}_{LO}}{q_c+a^{-1}_{LO}}\right| [n_B(T)]\right\} \\
& + \f{\a \hbar \o_{LO}}{a_{LO}} \Re \f{1}{ \pi z} \mathrm{ln} \left(\f{q_c-z}{q_c+z} \f{q_{\textrm{mesh}}+z}{q_{\textrm{mesh}}-z} \right) [2n_B(T)+1]
\label{correction_cond}
\end{aligned}
\end{flalign}

\ni Note that this is just the difference between equations \eqref{frohlich}, evaluated between 0 and $q_c$, and \eqref{adiabatic}, evaluated between $q_{\textrm{mesh}}$ and $q_c$. The plot of the adiabatic calculation with $\d=0.1$eV and the correction is included in Fig. \ref{adiabatic_corrected} in the main text.

The effective mass varies between $0.157 m_e$ and $0.175 m_e$ when taking different radii up to $0.067$ $2 \pi/a$. The difference of the Fr\"{o}hlich contribution for these two effective masses in the correction is only $0.6$ meV or less for all temperatures. So the change of the effective mass with $k$ causes negligible errors in our method.

\subsubsection{Valence band}

The expression we use for the correction is
\begin{flalign}
\begin{aligned}
\Delta & (E_{\mathbf{k}v}-\vare_{\mathbf{k}v})_{\mathbf{k}=0} = + \f{e^2}{4 \pi \tilde{\e_0} \Omega_0}\f{2 \pi  \hbar \o_{LO}}{\e^\ast} (I_{\textrm{Fr}}-I_{\textrm{Ad}}), \textrm{where}\\
& I_{\textrm{Fr}} = \f{\Omega_0}{(2 \pi)^3} \sum_{s=1}^3 \int_0^{q_c} \f{d^3q}{q^2} |\langle n_s(\mathbf{q}) | n \rangle |^2 \Re  \left[ \f{n_B(T)+1}{\vare_{\Gamma v}-\vare_{\mathbf{\mathbf{q}}n_s}+\hbar \o_{LO}} + \f{n_B(T)}{\vare_{\Gamma v}-\vare_{\mathbf{q}n_s}-\hbar \o_{LO} + i \Delta} \right] \\
& I_{\textrm{Ad}} = \f{\Omega_0}{(2 \pi)^3} \sum_{s=1}^3 \int_{q_{\textrm{mesh}}}^{q_c} \f{d^3q}{q^2} |\langle n_s(\mathbf{q}) | n \rangle |^2
\Re \left[ \f{2 n_B(T)+1}{\vare_{\Gamma v}-\vare_{\mathbf{\mathbf{q}}n_s}+i \d} \right]
\end{aligned}
\label{general}
\end{flalign}
\ni Here, $I_{\mathrm{Fr}}$ corresponds to Eq. \eqref{frohlich_valence} and $I_{\textrm{Ad}}$ is the corresponding adiabatic equation (these expressions are analogous to Eq. \eqref{frohlich} and \eqref{adiabatic} for the conduction band). As a reminder, we use $\d=0.1$ eV, the value that was used in the DFT calculation, and $\Delta=0.001$ eV to calculate the principal value adequately. The values of $\vare_{\mathbf{q}n_s}$ and $n_s(\mathbf{q})$ come from diagonalizing the matrix in Eq. \eqref{kp_method}, so Eq. \eqref{frohlich_valence} can be readily calculated, albeit not having an analytic expression as for the conduction band.

We study the convergence radius $q_c$ in the same way as we did for the conduction band. At $T=0$, we obtain $q_c= 2.5 q_{\textrm{mesh}}$ (for larger radii, the correction changes by less than 1 meV). At $T=1000 K$, however, we observe a difference of 6 meV between the correction at $q=2.5 q_{\textrm{mesh}}$ and the convergence radius $q_c=6.3 q_{\textrm{mesh}}$. What occurs is that the absorption term in the Fr\"ohlich integral changes more with the radius of integration than the emission term in $I_{\textrm{Fr}}$, and $I_{\textrm{Ad}}$. While the absorption term does not contribute at $T=0$ because it is suppressed by $n_B(T)$, it does at higher temperatures.

\begin{figure}
\centering
\includegraphics[width=0.8\textwidth]{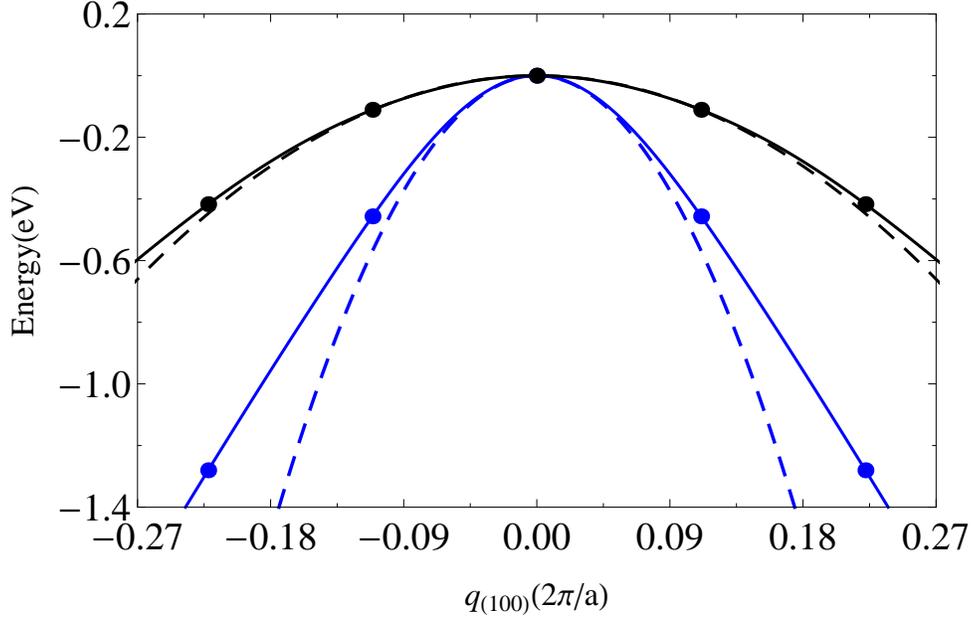}
\caption{Plot of the heavy and light holes in the (100) direction (full lines) with their corresponding effective mass  fit (dashed lines). The dots indicate the sampling points in the \textit{ab-initio} calculation.
\label{valence_bands}}
\end{figure}

From Fig. \ref{valence_bands} we  see that the effective mass approximation is accurate up to around $q_{m^\ast}=2.5 q_{\textrm{mesh}}=0.13$ $2\pi/a$ for the heavier masses, the same value we found for $q_c$. For the lighter mass, the effective mass approximation breaks down for a smaller $q$, but the convergence radius is much smaller (as for the conduction band). Varying the effective mass of the light hole, we can see that the error introduced is less than 1 meV (assuming a contribution of one third for each band; see the following paragraph). Therefore, our method is accurate for the valence band for temperatures below $500K$, and less accurate for higher temperatures.


In the isotropic case, it is shown in \cite{Trebin} how the renormalization is an average of the light and heavy holes at \textbf{k}=0. We can average the effective mass of each band over a sphere using Eq. \eqref{kp_method}. We obtain $m_{1,\textrm{av}}^\ast=0.14m_e$, $m_{2,\textrm{av}}^\ast=0.94m_e$ and $m_{3,\textrm{av}}^\ast=1.72m_e$. Assuming each band is isotropic, we can calculate the renormalization by using the standard Fr\"ohlich result Eq. \eqref{frohlich} for each band and then averaging over the bands. Integrating from 0 to $q_F$, with $0<q_F<6.3q_{\textrm{mesh}}$, the renormalization differs from Eq. \eqref{frohlich_valence} by less than 1 meV  at $T=0$. At $T=1000$ K, they differ by 5 meV or less, depending on the value of $q_F$. Therefore, at $T=0$ the renormalization can be just calculated by averaging over the Fr\"ohlich contribution of the average effective masses. At higher temperatures, using averaged isotropic masses is less accurate.

\subsection{Piezopolaron}

In a piezoelectric material, a strain induces a macroscopic electric field. If the strain is produced by a long-wavelength acoustic phonon, the coupled system of an electron and the acoustic phonon is known as the piezopolaron \cite{Hutson}. 
It turns out that, like the Fr\"ohlich polaron effect, the piezopolaron also causes a divergent intraband term at band 
extrema in the adiabatic approximation.  Adding an artificial $i \delta$ removes the divergence, but does not
correctly include the true non-adiabatic behavior, namely, part of the acoustic contribution to zero-point renormalization,
and a new low $T$ contribution scaling as $T^2$ with a positive coefficient (increasing the gap at low $T$).
This topic is covered in a separate paper in preparation \cite{PBAJPN}.  There we show that both the zero point
contribution and the high $T$ contribution are quite small, and the $T^2$ term only plays a dominant role at very
low $T$.  Therefore there is no need to add an analytic correction for piezo-effects
to the result obtained from adiabatic + $i\delta$ approximation.  To clarify a little, the formula for band
renormalization from intraband acoustic phonon processes, at a band extremum, is
\begin{equation}
[ E_{\mathrm{k}v} - \vare_{\mathrm{k}v} ]_{\mathrm{acoustic},\mathbf{k}=0}=\frac{1}{N}\sum_{\mathbf{q}j} | \langle\mathbf{q}| V_1(\mathbf{q}j)|\mathbf{0}\rangle|^2
\left[\frac{1+n_{\mathbf{q}j}} {-\hbar^2 q^2 /2m^\ast -\hbar v_j q}
+ \frac{n_{\mathbf{q}j}} {-\hbar^2 q^2 /2m^\ast +\hbar v_j q} \right], 
\label{eq:deltac}
\end{equation}
where $v_j$ is the velocity of sound.  This keeps the small $q$ part of the theory only.  The piezoelectric matrix element is
\be
\langle\mathbf{q}| V_1(\mathbf{q}j)|\mathbf{0}\rangle=- \f{e}{4 \pi \tilde{\vare}_0} \f{\mathbf{q} . \mathbf{e}_m . (\mathbf{q} \delta \mathbf{R})}{q^2 \e_{\infty}}
\ee

\ni where $\delta \mathbf{R}$ is the acoustic vibration or acoustic phonon amplitude and $\mathbf{e}_m$ is the electromechanical or piezoelectric tensor (see \cite{Cardona} for a derivation).  The acoustic phonon displacement factor $\delta R$
is $\sqrt{\hbar/M_{\rm tot}v_j q}$.  Therefore the squared matrix element 
$|\langle\mathbf{q}| V_1(\mathbf{q}s)|\mathbf{0}\rangle|^2$ behaves as $1/q$.
The adiabatic approximation replaces the factor $[ \ ]$ by the approximation $-[(1+2n_{\mathbf{q}j})/(\hbar^2 q^2/2m^\ast)]$.
Therefore, at low $T$, the sum over $\mathbf{q}$ becomes, at small $q$,
$-\int dq q^2 (1/q)[(1+2k_B T/\hbar v_j q)/(\hbar^2 q^2 /2m^\ast)]$.  This is valid for when the acoustic phonon energy 
$\hbar v_j q$ is smaller than $k_B T$.  The zero-point part diverges logarithmically, and the thermal part as $1/q$.  
The correct non-adiabatic version of this is
$\int dq q^2 (1/q)[-1/\hbar v_j q + (2k_B T/\hbar v_j q)(\hbar^2 q^2 /2m^\ast)/(\hbar v_j q)^2]$.  Both zero-point and thermal parts
converge as $\int dq $.   It turns out that the difference between the true non-adiabatic
contribution and the artificially converged adiabatic part (adding $+i\delta$ in the denominator) is small,
except for a small (but interesting) non-adiabatic $T^2$ term at very low $T$ which has little effect at higher $T$.


\begin{thebibliography}{100}

\bibitem{Olguin}  D. Olgu\'in, A. Cantarero and M. Cardona,
Temperature and Isotopic Mass Dependence of the Direct Band Gap in Semiconductors: LCAO Calculations 
Phys. Status Sol. {\bf 220}, 33 (2000).
\bibitem{Thewalt} M. Cardona and M. L. W. Thewalt, Isotope effects on the optical spectra of semiconductors,
Rev. Mod. Phys. {\bf 77}, 1173 (2005).
\bibitem{Franck} J. Franck and E. G. Dymond, Elementary processes of photochemical reactions, 
Trans. Faraday Soc. {\bf 21}, 536 (1926).
\bibitem{Condon} E. U. Condon, Nuclear Motions Associated with Electron Transitions in Diatomic Molecules, 
Phys. Rev. {\bf 32}, 858 (1928).
\bibitem{DeVreese} G. DeFilippis, V. Cautadella, A. S. Mishchenko, C. A. Perroni, and J. T. Devreese,
Validity of the Franck-Condon Principle in the Optical Spectroscopy: Optical Conductivity of the Fr\"ohlich Polaron,
Phys. Rev. Lett. {\bf 96}, 136405 (2006).
\bibitem{Berry} R. S. Berry, S. A. Rice, and J. Ross, {\it Physical Chemistry}, 2nd Ed. (Oxford, New York,
2000), pp.198-199.
\bibitem{Hui} P. B. Allen and J. C. K. Hui, Thermodynamics of solids: Corrections from electron-phonon
interactions, Z. Phys. B: Cond. Mat. {\bf 37}, 33 (1980).
\bibitem{Brooks} H. Brooks, in {\it Advances in Electronics and Electron Physics}, L. Marton and C. Marton, Eds. (Academic
Press, NY, 1955), v. 7, pp. 121-124.
\bibitem{Giustino_Louie_Cohen} F. Giustino, S. G. Louie, and M. L. Cohen, Electron-Phonon Renormalization of the Direct Band Gap of Diamond, Phys. Rev. Lett. \textbf{105}, 265501 (2010).
\bibitem{Gibbs} Z. M. Gibbs, H. Kim, H. Wang, R. L. White, F. Drymiotis, M. Kaviany, and G. J. Snyde, Temperature dependent band gap in PbX (X = S, Se, Te), Appl. Phys. Lett. \textbf{103}, 262109 (2013).
\bibitem{Ponce_Gonze} S. Ponc\'e, Y. Gillet, J. L. Janssen, A. Marini, M. Verstraete, and X. Gonze, Temperature dependence of the electronic structure of semiconductors and
insulators, J. Chem. Phys. \textbf{143}, 102813 (2015).
\bibitem{Ponce_Gonze2} S. Ponc\'{e}, G. Antonius, P. Boulanger, E. Cannuccia, A. Marini, M. C\^ot\'{e}, X. Gonze, Temperature dependence of electronic eigenenergies in the adiabatic harmonic approximation, Comp. Mater. Sci. \textbf{83}, 341 (2014).
\bibitem{Antonius_Gonze} G. Antonius, S. Ponc\'e, P. Boulanger, M. C\^ot\'e, and X. Gonze, Many-Body Effects on the Zero-Point Renormalization of the Band Structure, Phys. Rev. Lett. \textbf{112}, 215501 (2014).
\bibitem{metal1} S. M. Story, J. J. Kas, F. D. Vila, M. J. Verstraete and J. J. Rehr, Cumulant expansion for phonon contributions to the electron spectral function, Phys. Rev. B \textbf{90}, 195135 (2014).
\bibitem{metal2} M. J. Verstraete, Ab initio calculation of spin-dependent electron-phonon coupling in iron and
cobalt, J. Phys.: Condens. Matter \textbf{25}, 136001 (2013). 
\bibitem{Brown} A. M. Brown, R. Sundararaman, P. Narang, W. A. Goddard III and H. A. Atwater, Ab initio phonon coupling and optical response of hot electrons in plasmonic metals, arXiv:1602.00625v1.
\bibitem{Manjon} F. J. Manj\'{o}n, M. A. Hern\'{a}ndez-Fenollosa, B. Mar\'{i}, S. F. Li, C. D. Poweleit, A. Bell, J. Men\'{e}ndez, and M. Cardona, Effect of N isotopic mass on the photoluminescence and cathodoluminescence spectra of gallium nitride, Phys. J. B \textbf{40}, 453 (2004).
\bibitem{Cardona_Kremer} M. Cardona and R. K. Kremer, Temperature dependence of the electronic gaps of semiconductors, Thin Solid Films \textbf{571}, 680 (2011).
\bibitem{Bhosale} J. Bhosale and A. K. Ramdas, A. Burger, A. Munoz, A. H. Romero, M. Cardona, R. Lauck, and R. K. Kremer, Temperature dependence of band gaps in semiconductors: Electron-phonon interaction, Phys. Rev. B \textbf{86}, 195208 (2012).
\bibitem{Allen_Heine} P. B. Allen and V. Heine, Theory of the temperature dependence of electronic band structures, J. of Phys. C: Solid State Phys. \textbf{9}, 2305 (1976).
\bibitem{Allen_Cardona} P. B. Allen and M. Cardona, Theory of the temperature dependence of the direct gap of germanium, Phys. Rev. B \textbf{23}, 1495 (1981).
\bibitem{Slack} G. A. Slack, L. J. Schowalter, D. Morelli, and J. A. Freitas, J. Cryst. Growth \textbf{246}, 287 (2002).
\bibitem{Nakamura} S. Nakamura, G. Fasol, S.J. Pearton \textit{The Blue Laser Diode: The Complete Story} (Springer-Verlag, Berlin, 2000).
\bibitem{Khan} M.A. Khan, M.S. Shur, Q.C. Chen, and J.N. Kuznia, Current/voltage characteristic collapse in AlGaN/GaN heterostructure insulated gate field effect transistors at high drain bias, Electron. Lett. \textbf{30}, 2175 (1994).
\bibitem{Mohammad} S. N. Mohammad, A. A. Salvador, and H. Morko\c{c}, Emerging gallium nitride based devices, Proc. IEEE \textbf{83}, 1306 (1995).
\bibitem{Chung} K. Chung, C. H. Lee, and G. C. Yi, Transferable GaN layers grown on ZnO-coated graphene layers for optoelectronic devices, Science \textbf{330}, 655 (2010).
\bibitem{Mullhauser} J. R. M\"ullh\"auser, Dissertation: \textit{Properties of Zincblende GaN and (In,Ga,Al)N Heterostructures grown by Molecular Beam Epitaxy}, Paul-Drude-Institut f\"ur Festk\"orperelektronik, 
Humboldt-Universit\"at zu Berlin (1998).
\bibitem{Strite} S. Strite and H. Morko\c{c}, GaN, AlN, and InN: a review, J. Vac. Sci. and Technol. B \textbf{10}, 1237 (1992).
\bibitem{Brandt} O. Brandt, H. Yang, H. Kostial and K. H. Ploog, High p-type conductivity in cubic GaN/GaAs (113)A by using Be as the acceptor and O as the codopant, Appl. Phys. Lett. \textbf{69}, 2707 (1996).
\bibitem{Lin} M. E. Lin, G. Xue, G. L. Zhou, J. E. Greene and H. Morko\c{c}, p-type zincblende GaN on GaAs substrates, Appl. Phys. Lett. \textbf{63}, 932 (1993).
\bibitem{Vogl} P. Vogl, Microscopic theory of electron-phonon interaction in insulators or semiconductors, Phys. Rev. B \textbf{13}, 694 (1976).
\bibitem{Frohlich} H. Fr\"{o}hlich, Electrons in lattice fields, Adv. Phys. \textbf{3}, 325 (1954).
\bibitem{Verdi_Giustino} C. Verdi and F. Giustino, Fr\"ohlich Electron-Phonon Vertex from First Principles, Phys. Rev. Lett. \textbf{115}, 176401 (2015).
\bibitem{Mauri} J. Sjakste, N. Vast, M. Calandra and F. Mauri, Wannier interpolation of the electron-phonon matrix elements in polar semiconductors: Polar-optical coupling in GaAs, Phys. Rev B \textbf{92}, 054307 (2015).
\bibitem{Callaway} J. Callaway, \textit{Quantum Theory of the Solid State} (Academic Press, New York and London, 1974). 
\bibitem{Abinit1}  X. Gonze, B. Amadon, P. M. Anglade, J.-M. 
Beuken, F. Bottin, P. Boulanger, F. Bruneval, D. Caliste, R. Caracas, M. C\^ot\'e, T. Deutsch, L. Genovese, Ph. Ghosez, M. Giantomassi, S. Goedecker, D. Hamann, P. Hermet, F. Jollet, G. Jomard, S. Leroux, M. Mancini, S. Mazevet, M. J. T. Oliveira, G. Onida, Y. Pouillon, T. Rangel, G.-M. Rignanese, D. Sangalli, R. Shaltaf, M. Torrent, M.J. Verstraete, G. Zorah, J.W. Zwanziger, ABINIT: First-principles approach to material and nanosystem properties,  Comput. Phys. Commun. \textbf{180}, 2582 (2009). 
\bibitem{Abinit2} X. Gonze, J. M. Beuken, R. Caracas, F. Detraux, M. Fuchs, G. M. Rignanese, L. Sindic, M. Verstraete, G. Zerah, F. Jollet, M. Torrent, A. Roy, M. Mikami, Ph. Ghosez, J. Y. Raty, D. C. Allan, First-principles computation of material properties: the ABINIT software project, Comp. Mater. Sci. \textbf{25}, 478 (2002).
\bibitem{Perdew_Wang} J. P. Perdew and Y. Wang, Accurate and simple analytic representation of the electron-gas correlation energy, Phys. Rev. B \textbf{45}, 13244 (1992).
\bibitem{fhi98PP} M. Fuchs and M. Scheffler, Ab initio pseudopotentials for electronic structure calculations of poly-atomic systems using density-functional theory, Comp. Phys. Commun. {\bf 119}, 67 (1999).
\bibitem{MP} H. J. Monkhorst and J. D. Pack, Special points for Brillouin-zone integrations, Phys. Rev. B \textbf{13}, 5188 (1976).
\bibitem{Feneberg} M. Feneberg, M. R\"oppischer, C. Cobet, N. Esser, J. Sch\"ormann, T. Schupp, D. J. As, F. H\"orich, J. Bl\"asing, A. Krost, and R. Goldhahn, Optical properties of cubic GaN from 1 to 20 eV, Phys. Rev. B \textbf{85}, 155207 (2012).
\bibitem{Bougrov} Bougrov et al, in {\it Properties of Advanced Semiconductor Materials: GaN, AlN, InN, BN, SiC, SiGe}, edited by. M. E. Levinshtein, S. L. Rumyantsev, and M. S. Shur (Wiley, New York, 2001), pp. 1-30.
\bibitem{Baldereschi} A. Baldereschi and N. O. Lipari, Spherical Model of Shallow Acceptor States in Semiconductors, Phys. Rev. B \textbf{8}, 2697 (1973).
\bibitem{Trebin} H.-R. Trebin and U. R\"{o}ssler, Polarons in the Degenerate-Band Case, Phys. Stat. Sol. B \textbf{70}, 717 (1975).
\bibitem{Hutson} A. R. Hutson, Piezoelectric scattering and phonon drag in ZnO and CdS, J. Appl. Phys. \textbf{32}, 2287 (1961).
\bibitem{PBAJPN} P. B. Allen and J. P. Nery, arXiv:1605.01980.
\bibitem{Cardona} P. Yu and M. Cardona, \textit{Fundamentals of Semiconductors} (Springer-Verlag, Berlin, 1996), p. 122.

\end{thebibliography}
\end{document}